\documentclass{article}%
\usepackage{amssymb}
\usepackage{amsfonts}
\usepackage{amsmath}
\usepackage[numbers,sort&compress]{natbib}
\usepackage{graphicx}%
\providecommand{\U}[1]{\protect\rule{.1in}{.1in}}
\newtheorem{theorem}{Theorem}

\newtheorem{corollary}[theorem]{Corollary}

\newtheorem{example}{Example}

\newenvironment{proof}[1][Proof]{\noindent\textbf{#1.} }{\ \rule{0.5em}{0.5em}}
\numberwithin{equation}{section}
\begin{document}

\title{Miura maps for St\"{a}ckel systems}
\author{Krzysztof Marciniak\\Department of Science and Technology \\Campus Norrk\"{o}ping, Link\"{o}ping University\\601-74 Norrk\"{o}ping, Sweden\\\texttt{krzma@itn.liu.se}
\and Maciej B\l aszak\\Faculty of Physics, Department of of Mathematical Physics and Computer Modelling,\\A. Mickiewicz University, 61-614 Pozna\'{n}, Poland\\\texttt{blaszakm@amu.edu.pl}}
\maketitle

\begin{abstract}
We introduce the concept of Miura maps between parameter-dependent algebraic
curves of hyper-elliptic type. These Miura maps induce Miura maps between
St\"{a}ckel systems defined (on the extended phase space) by the considered
algebraic curves. This construction yields a new way of generating
multi-Hamiltonian representations for St\"{a}ckel systems.

\end{abstract}

\section{Introduction}

St\"{a}ckel systems are Hamiltonian systems that are generated by separation
relations affine in Hamiltonians. They are Liouville integrable and separable,
hence integrable in quadratures.

An important class of St\"{a}ckel systems is obtained from algebraic curves of
hyperelliptic-type on a plane (see \cite{blasz2019} and the literature
therein). Such St\"{a}ckel systems are bi-Hamiltonian on the extended phase
space \cite{blasz2009}. A subclass of this class are St\"{a}ckel systems that
are of Benenti type \cite{Ben1,Ben2} and it is these systems (or rather their
multi-parameter extensions, see below) that we will consider in this article.
To shorten the terminology, throughout the article we will simply call these
systems St\"{a}ckel systems.

In this article we introduce a new concept of Miura maps between
parameter-dependent algebraic curves of hyper-elliptic type on the plane, each
of them defining a parameter-dependent St\"{a}ckel system, i.e. a St\"{a}ckel
system in an extended phase space. These maps in turn generate Miura maps
between the corresponding St\"{a}ckel systems themselves in the following
sense: we show that each pair of these St\"{a}ckel systems, when considered on
the phase space extended by their respective parameters (that play the role of
Casimir variables in the extended phase space) can be identified by a
non-canonical map \cite{ant87}, being a finite-dimensional counterpart of
Miura maps known from the theory of integrable nonlinear field systems (see
for example \cite{M,M1,M2}). Further, with the help of these
finite-dimensional Miura maps, we construct multi-Hamiltonian representation
of considered St\"{a}ckel systems in all coordinate sets defined by our
equivalent multi-parameter separation curves.

St\"{a}ckel systems generated by different separation curves can sometimes be
identified using an appropriately chosen St\"{a}ckel transform
\cite{serg2008,SAPM2012}. The Miura maps that we construct in this article do
not constitute any St\"{a}ckel transform as St\"{a}ckel transform in general
requires transforming times (the independent variables)\ as well. Our
construction can therefore not be reduced to any St\"{a}ckel transform.

\section{St\"{a}ckel systems}

Consider the following algebraic (hyperelliptic-type) curve in the
$(\lambda,\mu)$-plane
\begin{equation}
\sigma(\lambda)+\sum_{r=1}^{n}h_{r}\lambda^{n-r}=f(\lambda)\mu^{2}
\label{2.1new}%
\end{equation}
(with $\lambda$ and $\mu$ $\in\mathbf{R}$) where $f$ and $\sigma$ are
arbitrary Laurent polynomials in $\lambda$. Taking $n$ copies of
(\ref{2.1new}) at points $(\lambda_{i},\mu_{i})$, $i=1,\dotsc,n$, we obtain a
system of $n$ linear equations (separation relations) for $h_{r}$:
\[
\sigma(\lambda_{i})+\sum_{r=1}^{n}h_{r}\lambda_{i}^{n-r}=f(\lambda_{i})\mu
_{i}^{2}\text{, \ \ }i=1,\ldots,n.
\]
Solving this system (by inverting the Vandermonde matrix $\lambda_{i}^{n-r}$)
yields $n$ functions (Hamiltonians)%
\begin{equation}
h_{r}=E_{r}+V_{r}^{(\sigma)},\quad r=1,\dotsc,n \label{2.2new}%
\end{equation}
on a $2n$-dimensional manifold (phase space) $M=\mathbf{R}^{2n}=T^{\ast}Q$
with $Q=\mathbf{R}^{n}$. The manifold $M$ is parametrized by the coordinates
$(\lambda_{1},\ldots,\lambda_{n},\mu_{1},\ldots\mu_{n})$ in such a way that
$\lambda_{i}$ are coordinates on $Q$ while $\mu_{i}$ are fiber coordinates in
$T^{\ast}Q$. The geodesic parts $E_{r}$ of (\ref{2.2new}) are given by
\[
E_{r}=\mu^{T}K_{r}G\mu\text{, \ \ }r=1,\ldots,n,\text{\ \ }%
\]
with $G\,$ treated as a contravariant metric tensor on the configurational
space $Q$, $K_{r}$ ($K_{1}=\operatorname{Id}$) are $(1,1)$-Killing tensors of
$G$ and $V_{r}^{(\sigma)}$ are separable potentials on $Q$. In the coordinates
$\lambda_{i}$ on $Q$ the geometric objects $G$, $K$ and $V_{r}^{(\sigma)}$ are
explicitly given by
\begin{equation}
G^{ij}=\frac{f(\lambda_{i})}{\Delta_{i}}\delta^{ij}\text{, \ }(K_{r})_{j}%
^{i}=-\frac{\partial\rho_{r}}{\partial\lambda_{i}}\delta_{j}^{i}\text{,
\ \ }V_{r}^{(\sigma)}=\sum_{j=1}^{n}\frac{\partial\rho_{r}}{\partial
\lambda_{j}}\frac{\sigma(\lambda_{j})}{\Delta_{j}} \label{2.4}%
\end{equation}
(no summation in the above formulas unless explicitly stated) where
$\Delta_{j}=%
{\textstyle\prod\nolimits_{k\neq j}}
(\lambda_{j}-\lambda_{k})$ and $\rho_{r}=(-1)^{r}s_{r}$ where $s_{r}$ are
elementary symmetric polynomials in $n$ variables $\lambda_{i}$. The
coordinates $\lambda_{i}$ are thus orthogonal coordinates for the metric $G$.

By construction, all the Hamiltonian functions $h_{r}$ are in involution
\begin{equation}
\{h_{r},h_{s}\}\equiv\pi(dh_{r},dh_{s}%
)=0,\ \ \ \ \ \ \ r,s=1,...,n\label{konwencja}%
\end{equation}
with respect to the Poisson bracket $\pi=%
{\textstyle\sum_{i=1}^{n}}
\frac{\partial}{\partial\lambda_{i}}\wedge\frac{\partial}{\partial\mu_{i}}$ on
$M$, i.e. the Hamiltonians (\ref{2.2new}) constitute a Liouville integrable
system. By construction they also separate in coordinates $(\lambda_{i}%
,\mu_{i})_{i=1,...,n}$ (in the sense of Hamilton-Jacobi theory). The
Hamiltonians (\ref{2.2new}) are known in literature as St\"{a}ckel
Hamiltonians of Benenti type. The Hamiltonians (\ref{2.2new}) yield the system
of commuting Hamiltonian flows on $M$:%
\begin{equation}
\frac{d\xi}{dt_{r}}=X_{r}=\pi dh_{r}\text{, \ \ \ \ }r=1,\ldots,n.\label{ds}%
\end{equation}
(where $\xi\in M$). The geometric properties of such systems were investigated
by Benenti et al. in \cite{Ben1,Ben2,Ben3,Ben4}; cf. also \cite{s2007}.

\section{Miura maps between St\"{a}ckel systems in the extended phase space}

In this section we consider $1\leq N\leq n$ specific extensions of the curve
(\ref{2.1new}) by $N$ additional real parameters each. Each such extended
curve defines a St\"{a}ckel system in the phase space $\mathcal{M=}%
\mathbf{R}^{2n+N}$ extended by these parameters treated as new variables. With
each curve we associate, in a very natural way, a degenerated Poisson tensor
and the extra parameters (extra variables) become Casimirs of this Poisson
tensors.

We show that these extended curves can be related by specific transformations
which we will call \emph{Miura maps}, since they induce finite dimensional
counterparts of Miura maps (known from soliton theory) in the extended phase
space $\mathbf{R}^{2n+N}$ between the corresponding St\"{a}ckel systems
generated by these curves. These Miura maps on $\mathcal{M=}\mathbf{R}^{2n+N}$
are non-canonical transformations. We thus prove that all the extended
St\"{a}ckel systems actually represent the same system, written in different
coordinates, connected by Miura maps. This in turn will lead to a
multi-Hamiltonian formulation of the considered St\"{a}ckel system.

Let us thus consider the curve (\ref{2.1new}) extended by $N$ additional real
parameters $c_{i}$ where $1\leq N\leq n$%
\begin{equation}
\sigma(\lambda)+c_{N}\lambda^{n+N-1}+\ldots+c_{1}\lambda^{n}+h_{1}%
\lambda^{n-1}+\ldots+h_{n}=f(\lambda)\mu^{2}.\label{ext}%
\end{equation}
It defines the parameter-dependent St\"{a}ckel system%
\begin{equation}
\frac{d\xi}{dt_{r}}=X_{r}\equiv\pi dh_{r}(\lambda,\mu,c)\text{, \  }%
r=1,\ldots,n\text{, \ \ }\pi=%
{\textstyle\sum_{i=1}^{n}}
\frac{\partial}{\partial\lambda_{i}}\wedge\frac{\partial}{\partial\mu_{i}%
}.\label{S0}%
\end{equation}
(with $\xi\in M$) Consider also, for any $s\in\left\{  1,\ldots,N\right\}  $,
the $N$-parameter curve%
\begin{equation}
\sigma(\bar{\lambda})\bar{\lambda}^{-s}+\bar{c}_{N}\bar{\lambda}%
^{n+N-s-1}+\ldots+\bar{c}_{s+1}\bar{\lambda}^{n}+\bar{h}_{1}\bar{\lambda
}^{n-1}+\ldots+\bar{h}_{n}+\bar{c}_{s}\bar{\lambda}^{-1}+\ldots+\bar{c}%
_{1}\bar{\lambda}^{-s}=f(\bar{\lambda})\bar{\lambda}^{s}\bar{\mu}%
^{2}.\label{ext2}%
\end{equation}
It defines the parameter-dependent St\"{a}ckel system%
\begin{equation}
\frac{d\bar{\xi}}{dt_{r}}=\bar{X}_{r}\equiv\bar{\pi}d\bar{h}_{r}%
(\bar{\lambda},\bar{\mu},\bar{c})\text{, \ \  }r=1,\ldots,n\text{, \ \ }%
\bar{\pi}=%
{\textstyle\sum_{i=1}^{n}}
\frac{\partial}{\partial\bar{\lambda}_{i}}\wedge\frac{\partial}{\partial
\bar{\mu}_{i}}\label{Ss}%
\end{equation}
(with $\bar{\xi}\in M$). Note that with each curve $s$ we
associate a degenerated Poisson tensor $\bar{\pi}$ that is
canonical in the variables
$(\bar{\lambda}_{i},\bar{\mu}_{i},\bar{c}_{j})_{i=1,...,n;\
j=1,...,N}$.

We will now attempt to find a map between the variables $(\lambda_{i},\mu
_{i},c_{j})_{i=1,...,n;\ j=1,...,N}$ and the variables $(\bar{\lambda}%
_{i},\bar{\mu}_{i},\bar{c}_{j})_{i=1,...,n;\ j=1,...,N}$ that maps the system
(\ref{S0}) to the system (\ref{Ss}). Consider thus the following map in
$\mathbf{R}^{2}$:%
\begin{equation}
\bar{\lambda}=\lambda,\text{ \ \ }\bar{\mu}=\lambda^{-s}\mu.\label{pmap}%
\end{equation}
This map transforms (algebraically) the curve (\ref{ext}) into the curve
(\ref{ext2}), provided that%
\begin{equation}%
\begin{array}
[c]{ll}%
\bar{c}_{i}=h_{n-i+1}, & i=1,\ldots,s\\
\bar{c}_{i}=c_{i}, & i=s+1,\ldots,N\\
\bar{h}_{i}=c_{s-i+1}, & i=1,\ldots,s\\
\bar{h}_{i}=h_{i-s}, & i=s+1,\ldots,n.
\end{array}
\label{map}%
\end{equation}
We will thus call the pair (\ref{pmap})-(\ref{map}) a \emph{Miura map between
curves} (\ref{ext})\ and (\ref{ext2}). The relations (\ref{map}) can be
inverted to%
\begin{equation}%
\begin{array}
[c]{ll}%
c_{i}=\bar{c}_{i}, & i=s+1,\ldots,N\\
c_{i}=\bar{h}_{s-i+1}, & i=1,\ldots,s\\
h_{i}=\bar{h}_{s+i}, & i=1,\ldots,n-s\\
h_{i}=\bar{c}_{n-i+1}, & i=n-s+1,\ldots,n.
\end{array}
\label{mapinv}%
\end{equation}
The Miura map (\ref{pmap})-(\ref{map}) induces the following \emph{Miura map
on the extended phase space} $M_{0\rightarrow s}:\mathbf{R}^{2n+N}%
\rightarrow\mathbf{R}^{2n+N}$%

\begin{equation}%
\begin{array}
[c]{ll}%
\bar{\lambda}_{i}=\lambda_{i}, & i=1,\ldots,n\\
\bar{\mu}_{i}=\lambda_{i}^{-s}\mu_{i}, & i=1,\ldots,n\\
\bar{c}_{i}=h_{n-i+1}(\lambda,\mu,c), & i=1,\ldots,s\\
\bar{c}_{i}=c_{i}, & i=s+1,\ldots,N,
\end{array}
\label{miura}%
\end{equation}
with the inverse $M_{0\rightarrow s}^{-1}:\mathbf{R}^{2n+N}\rightarrow
\mathbf{R}^{2n+N}$%
\begin{equation}%
\begin{array}
[c]{ll}%
\lambda_{i}=\bar{\lambda}_{i}, & i=1,\ldots,n\\
\mu_{i}=\bar{\lambda}_{i}^{s}\bar{\mu}_{i}, & i=1,\ldots,n\\
c_{i}=\bar{h}_{s-i+1}(\bar{\lambda},\bar{\mu},\bar{c}), & i=1,\ldots,s\\
c_{i}=\bar{c}_{i}, & i=s+1,\ldots,N.
\end{array}
\label{minv}%
\end{equation}

\begin{theorem}
\label{obvious}For any $s\in\{1,\ldots,N\}$ we have
\[
X_{r}=\bar{X}_{r},\text{ \ \ }r=1,\ldots,n.
\]

\end{theorem}

The proof of this theorem is in the Appendix.

Thus, all the Hamiltonian vector fields $X_{r}$ and $\bar{X}_{r}$ pairwise
coincide on $\mathcal{M}$. This means that all the St\"{a}ckel systems
(\ref{Ss}), generated by the $N$ curves (\ref{ext2}) (one for each value of
$s$ between $1$ and $N$), represent on the extended phase space $\mathcal{M=}%
\mathbf{R}^{2n+N}$ the same St\"{a}ckel system as (\ref{S0}), written in
different coordinates, connected by the corresponding invertible Miura map
(\ref{miura}).

By composing two appropriate Miura maps of type (\ref{miura}) we can now
easily relate St\"{a}ckel systems generated by two curves of type
(\ref{ext2}), i.e%
\begin{equation}
\sigma(\lambda)\lambda^{-r}+c_{N}\lambda^{n+N-r-1}+\ldots+c_{r+1}\lambda
^{n}+h_{1}\lambda^{n-1}+\ldots+h_{n}+c_{r}\lambda^{-1}+\ldots+c_{1}%
\lambda^{-r}=f(\lambda)\lambda^{r}\mu^{2}\label{kr}%
\end{equation}
and
\begin{equation}
\sigma(\bar{\lambda})\bar{\lambda}^{-s}+\bar{c}_{N}\bar{\lambda}%
^{n+N-s-1}+\ldots+\bar{c}_{s+1}\bar{\lambda}^{n}+\bar{h}_{1}\bar{\lambda
}^{n-1}+\ldots+\bar{h}_{n}+\bar{c}_{s}\bar{\lambda}^{-1}+\ldots+\bar{c}%
_{1}\bar{\lambda}^{-s}=f(\bar{\lambda})\bar{\lambda}^{s}\bar{\mu}%
^{2}\label{ks}%
\end{equation}
(with $0\leq r<s\leq N\leq n$), by a Miura map (note how the parameters
$c_{i}$ respectively $\bar{c}_{i}$ are distributed in both curves). Consider
thus the following map in the plane $\mathbf{R}^{2}$%
\begin{equation}
\bar{\lambda}=\lambda,\text{ \ }\bar{\mu}=\lambda^{r-s}\mu.\label{jola}%
\end{equation}
This map transforms the curve (\ref{kr}) into the curve (\ref{ks}), provided
that%
\begin{equation}%
\begin{array}
[c]{ll}%
\bar{c}_{i}=h_{n+r-i+1}, & i=r+1,\ldots,s\\
\bar{c}_{i}=c_{i}, & i=1,\ldots,r,s+1,\ldots,N\\
\bar{h}_{i}=c_{s-i+1}, & i=1,\ldots,s-r\\
\bar{h}_{i}=h_{i-(s-r)}, & i=s-r+1,\ldots,n.
\end{array}
\label{mrs}%
\end{equation}
The maps (\ref{jola})-(\ref{mrs}) constitute a Miura map between the curves
(\ref{kr})\ and (\ref{ks}). The relations (\ref{mrs}) can be inverted to%
\begin{equation}%
\begin{array}
[c]{ll}%
c_{i}=\bar{c}_{i}, & i=1,\ldots,r,s+1,\ldots,N\\
c_{i}=\bar{h}_{s-i+1}, & i=r+1,\ldots,s\\
h_{i}=\bar{h}_{s-r+i}, & i=1,\ldots,n-(s-r)\\
h_{i}=\bar{c}_{n+r-i+1}, & i=n-(s-r)+1,\ldots,n.
\end{array}
\label{mrsinv}%
\end{equation}
Note that these maps properly reduce to the maps (\ref{map}) and
(\ref{mapinv}), respectively, in the case that $r=0$.

The Miura map (\ref{jola})-(\ref{mrs}) induces the following Miura map
$M_{r\rightarrow s}$ on $\mathbf{R}^{2n+N}$%

\begin{equation}%
\begin{array}
[c]{ll}%
\bar{\lambda}_{i}=\lambda_{i}, & i=1,\ldots,n\\
\bar{\mu}_{i}=\lambda_{i}^{r-s}\mu_{i}, & i=1,\ldots,n\\
\bar{c}_{i}=h_{n+r-i+1}(\lambda,\mu,c) & i=r+1,\ldots,s\\
\bar{c}_{i}=c_{i}, & i=1,\ldots,r,s+1,\ldots,N
\end{array}
\label{Mrs}%
\end{equation}
that is nothing else than the composition $M_{r\rightarrow
s}=M_{0\rightarrow s}\circ M_{0\rightarrow r}^{-1}$. This map has
the inverse $M_{r\rightarrow
s}^{-1}$%
\begin{equation}%
\begin{array}
[c]{ll}%
\lambda_{i}=\bar{\lambda}_{i}, & i=1,\ldots,n\\
\mu_{i}=\bar{\lambda}_{i}^{s-r}\bar{\mu}_{i}, & i=1,\ldots,n\\
c_{i}=\bar{h}_{s-i+1}(\bar{\lambda},\bar{\mu},\bar{c}), & i=r+1,\ldots,s\\
c_{i}=\bar{c}_{i}, & i=1,\ldots,r,s+1,\ldots,N
\end{array}
\label{Mrsinv}%
\end{equation}
that is the composition $M_{r\rightarrow s}^{-1}=M_{0\rightarrow r}\circ
M_{0\rightarrow s}^{-1}$. Thus, again, the systems (\ref{kr}) and (\ref{ks})
represent in fact the same system expressed in different separable coordinate
systems connected by the Miura map (\ref{Mrs}). Note also that the Miura maps
(\ref{Mrs}) and (\ref{Mrsinv})\ properly reduce to the Miura maps
(\ref{miura}) and (\ref{minv}), respectively, for $r=0$.

\section{Miura maps and multi-Hamiltonian structure of St\"{a}ckel systems}

In the previous chapter we introduced $N+1\,\ $co-rank $N$ Poisson operators
$\pi_{0},\ldots,\pi_{N}$ on $\mathbf{R}^{2n+N}$ defined as follows. The
operator $\pi_{0}$ is canonical with respect to the separation variables
$(\lambda_{i},\mu_{i},c_{j})_{i=1,...,n;\ j=1,...,N}$ of the curve (\ref{ext})
\
\[
\pi_{0}=\sum_{i=1}^{n}\frac{\partial}{\partial\lambda_{i}}\wedge\frac
{\partial}{\partial\mu_{i}}%
\]
and for a given $s\in\left\{  1,\ldots,N\right\}  $ by $\pi_{s}$ we denote the
operator
\[
\pi_{s}=\sum_{i=1}^{n}\frac{\partial}{\partial\bar{\lambda}_{i}}\wedge
\frac{\partial}{\partial\bar{\mu}_{i}},
\]
that is canonical with respect to the separation variables $(\bar{\lambda}%
_{i},\bar{\mu}_{i},\bar{c}_{j})_{i=1,...,n;\ j=1,...,N}$ of the curve
(\ref{ext2}) with fixed $s$. Both sets of variables are related by the Miura
map (\ref{miura}).

In this section we use Miura maps, introduced in the previous section, to
find, in a novel way,  multi-Hamiltonian structure of the considered
St\"{a}ckel system on $\mathcal{M}$. We  start by formulating a theorem that
clarifies how all the operators $\pi_{r}$, with $r\in\{0,\ldots,N\}$, can be
expressed by the base vector fields of the coordinates $(\bar{\lambda}%
,\bar{\mu},\bar{c})$\ (those associated with the curve $s$ i.e. the
coordinates in which $\pi_{s}$ is canonical).

\begin{theorem}
\label{trpi}For $r\leq s$%
\begin{equation}
\pi_{r}=\sum_{i=1}^{n}\bar{\lambda}_{i}^{r-s}\frac{\partial}{\partial
\bar{\lambda}_{i}}\wedge\frac{\partial}{\partial\bar{\mu}_{i}}\text{ }%
+\sum_{j=1}^{s-r}X_{n-j+1}\wedge\frac{\partial}{\partial\bar{c}_{r+j}}
\label{1}%
\end{equation}
while for $r\geq s$%
\begin{equation}
\pi_{r}=\sum_{i=1}^{n}\bar{\lambda}_{i}^{r-s}\frac{\partial}{\partial
\bar{\lambda}_{i}}\wedge\frac{\partial}{\partial\bar{\mu}_{i}}\text{ }%
+\sum_{j=1}^{r-s}X_{j}\wedge\frac{\partial}{\partial\bar{c}_{r-j+1}}.
\label{2}%
\end{equation}

\end{theorem}

Note that for $r=s$ the formulas (\ref{1}) and (\ref{2}) coincide and yield
then the same (canonical in $(\bar{\lambda},\bar{\mu},\bar{c})$) Poisson
operator $\pi_{s}=%
{\textstyle\sum_{i=1}^{n}}
\frac{\partial}{\partial\bar{\lambda}_{i}}\wedge\frac{\partial}{\partial
\bar{\mu}_{i}}$.

Theorem \ref{trpi} is equivalent to the statement that the matrix
representation $\pi_{r}(\bar{\lambda},\bar{\mu},\bar{c})$ of the Poisson
operator $\pi_{r}$ in the variables $(\bar{\lambda},\bar{\mu},\bar{c})$
(associated with the curve $s$) is: for $r<s$%
\begin{equation}
\pi_{r}(\bar{\lambda},\bar{\mu},\bar{c})=\left(
\begin{array}
[c]{cc}%
\begin{array}
[c]{ll}%
\ \ 0 & \Lambda^{r-s}\\
-\Lambda^{r-s} & 0
\end{array}
& \overset{r}{\overbrace{0...0}}\text{ \ }X_{n}\ \cdots\ X_{n+r-s+1}\text{
}\overset{N-s}{\overbrace{0...0}}\\
\ast & \ \ \ \ \ \ \ \ \ \ \ \ \ \ \ \ \ \
\end{array}
\right)  \label{pirs1}%
\end{equation}
where $\Lambda=\mathrm{diag}(\bar{\lambda}_{1},\ldots,\bar{\lambda}_{n}),$
while for $r>s$%
\begin{equation}
\pi_{r}(\bar{\lambda},\bar{\mu},\bar{c})=\left(
\begin{array}
[c]{cc}%
\begin{array}
[c]{ll}%
\ \ 0 & \Lambda^{r-s}\\
-\Lambda^{r-s} & 0
\end{array}
& \overset{s}{\overbrace{0...0}}\text{ \ }X_{r-s}\ \cdots\ X_{1}\text{
}\overset{N-r}{\overbrace{0...0}}\\
\ast & \ \ \ \ \ \ \ \ \ \ \ \ \ \ \ \ \ \
\end{array}
\right)  \label{pirs2}%
\end{equation}
(where $\ast$ denote the elements that make the matrix $\bar{\pi}_{r}$
antisymmetric). In both matrices (\ref{pirs1}) and (\ref{pirs2}) above by
$X_{i}$ we denote the columns consisting of components of the vector field in
coordinates $(\bar{\lambda},\bar{\mu},\bar{c})$. Note that for $r=s$ both
matrices (\ref{pirs1})\ and (\ref{pirs2}) reduce to the same matrix of the
canonical - in the variables $(\bar{\lambda},\bar{\mu},\bar{c})$ -\ operator
$\pi_{s}=%
{\textstyle\sum_{i=1}^{n}}
\frac{\partial}{\partial\bar{\lambda}_{i}}\wedge\frac{\partial}{\partial
\bar{\mu}_{i}}$, with $N$ Casimir functions $\bar{c}_{1},\ldots,\bar{c}_{N}$.

\begin{proof}
Let us first prove (\ref{pirs1}). To prove this we will use the direct Miura
map (\ref{Mrs}) between the variables associated with curve $r$ (denoted there
and in this proof by $(\lambda,\mu,c)$) and the variables $(\bar{\lambda}%
,\bar{\mu},\bar{c})$ associated with our fixed curve $s$. The upper left
$2n\times2n\,\ $block in (\ref{pirs1}) is easily proven by a direct
calculation from the first part of (\ref{Mrs}). The column $2n+i,$
$i=1,\ldots,N,$ of $\pi_{r}$ in (\ref{pirs1}) consists of components of the
vector fields $\pi_{r}d\bar{c}_{i}$ in coordinates $(\bar{\lambda},\bar{\mu
},\bar{c})$. Due to (\ref{Mrs}), $\pi_{r}d\bar{c}_{i}=\pi_{r}dc_{i}=0$ for
$i=1,\ldots,r,s+1,\ldots,N$, since $c_{i}$ are in this notation Casimirs of
$\pi_{r}$. Thus all the columns $2n+1,\ldots,2n+r,2n+s+1,\ldots2n+N$ of
$\pi_{r}$ contain only zeros. Further, due to (\ref{Mrs}) and to Theorem
\ref{obvious},
\[
\pi_{r}d\bar{c}_{i}=\pi_{r}dh_{n+r-i+1}(\lambda,\mu,c)=X_{n+r-i+1}\text{ for
}i=r+1,\ldots,s
\]
which yields exactly the columns $2n+r+1,\ldots,2n+s$ as in (\ref{pirs1}). The
formula (\ref{pirs2}) can be proved in an analogous way but there we have to
use the inverse Miura map (\ref{Mrsinv})\ with the roles of $r$ and $s$
interchanged, as the formula (\ref{Mrs}) is formulated for the case $r<s$ only.
\end{proof}

From the above considerations it also follows that $N$ Casimirs of $\pi_{r}$,
when written in the variables $(\bar{\lambda},\bar{\mu},\bar{c})$ associated
with our fixed curve $s$, are: for $r<s$%
\[%
\begin{array}
[c]{l}%
\bar{c}_{i}\text{,\ \ \ \ \ \ \ \ \ \ \ \ \ \ \ \ \ \ \ \ }i=1,\ldots
,r,s+1,\ldots,N,\\
\bar{h}_{s+1-i}(\bar{\lambda},\bar{\mu},\bar{c})\text{,\ \ \ \ }i=r+1,\ldots,s
\end{array}
\]
and for $r>s$%
\[%
\begin{array}
[c]{l}%
\bar{c}_{i}\text{,\ \ \ \ \ \ \ \ \ \ \ \ \ \ \ \ \ \ \ \ }i=1,\ldots
,s,r+1,\ldots,N,\\
\bar{h}_{n-r+i}(\bar{\lambda},\bar{\mu},\bar{c})\text{,\ \ \ \ }%
i=s+1,\ldots,r.
\end{array}
\]

\begin{corollary}
All the Poisson operators $\pi_{s}$, $s=0,...,N$ are pairwise compatible. It
follows from Theorem (\ref{trpi}) and from the fact, that $\pi_{0}$ and
$\pi_{s}$ are compatible, as it has been proved in \cite{blasz1999} (in the
special case $N=n$ but the proof in \cite{blasz1999} can easily be generalized
to the case $N<n$).
\end{corollary}

Denoting
\[
h_{j}\equiv c_{1-j}\text{ \ for }j=0,-1,\ldots,-N+1,
\]
due to Theorem \ref{obvious} and (\ref{map}), we obtain
\begin{align*}
X_{k} &  =\pi_{0}dh_{k}=\pi_{r}d\bar{h}_{k}=\pi_{r}dc_{r-k+1}=\pi_{r}%
dh_{k-r},\text{ \ \ }k=1,\ldots r,\\
X_{k} &  =\pi_{0}dh_{k}=\pi_{r}d\bar{h}_{k}=\pi_{r}dh_{k-r},\text{
\ }k=r+1,\ldots n.
\end{align*}
The vector fields $X_{k},$ $k=1,...,n$ are thus \thinspace$(N+1)$-Hamiltonian:%
\begin{equation}
X_{k}=\pi_{0}dh_{k}=\pi_{1}dh_{k-1}=\cdots=\pi_{N}dh_{k-N},\text{
\ \ }k=1,\ldots,N\label{bH}%
\end{equation}
and therefore generate $\binom{N+1}{2}$ bi-Hamiltonian chains
\begin{equation}%
\begin{array}
[c]{l}%
\pi_{k}dh_{-k}\ \ \ =0\\
\pi_{k}dh_{-k+1}=X_{1}=\pi_{r}dh_{-r+1}\\
\ \ \ \ \ \ \ \ \ \ \ \ \ \ \ \ \vdots\\
\pi_{k}dh_{-k+i}=X_{i}=\pi_{r}dh_{-r+i}\\
\ \ \ \ \ \ \ \ \ \ \ \ \ \ \ \ \vdots\\
\pi_{k}dh_{-k+n}=X_{n}=\pi_{r}dh_{-r+n}\\
\ \ \ \ \ \ \ \ \ \ \ \ \ \ \ \ \ \ \ \ 0=\pi_{r}dh_{-r+n+1}%
\end{array}
\label{bh}%
\end{equation}
for $0\leq k<r\leq N$, where $h_{i}$ are defined by the $s=0$ curve (\ref{ext}).

The multi-Hamiltonian representations (\ref{bh}) of St\"{a}ckel systems,
generated by separation curves (\ref{ext}), was proved for the first time in
\cite{blasz1999} by explicit calculations.

\begin{example}
Let us consider the St\"{a}ckel system generated by the curve (\ref{ext}) with
$n=2$, $N=2$, $\sigma(\lambda)=\lambda^{4}$, $f(\lambda)=1$:
\begin{equation}
\lambda^{4}+c_{2}\lambda^{3}+c_{1}\lambda^{2}+h_{1}\lambda+h_{2}=\mu^{2}.
\label{k1}%
\end{equation}
It can be shown that this system is the third stationary system in the
Dispersive Water Wave (DWW) hierarchy, see \cite{BMK}. For $s=0$ (that is in
the variables $(\lambda,\mu,c)$ associated with the curve (\ref{k1})) the
matrix representations of the operators $\pi_{r}$ on $\mathcal{M}=R^{6}$ are
given by the formulas (\ref{pirs1}) and (\ref{pirs2})\ that specify to%
\[
\pi_{0}(\lambda,\mu,c)=\left(
\begin{array}
[c]{cc}%
\begin{array}
[c]{ll}%
\ \ 0 & I\\
-I & 0
\end{array}
& 0\ \ \ 0\\
\ast &
\end{array}
\right)  ,\ \ \ \pi_{1}(\lambda,\mu,c)=\left(
\begin{array}
[c]{cc}%
\begin{array}
[c]{ll}%
\ \ 0 & \Lambda\\
-\Lambda & 0
\end{array}
& X_{1}\ \ \ 0\\
\ast &
\end{array}
\right)  ,
\]%
\[
\pi_{2}(\lambda,\mu,c)=\left(
\begin{array}
[c]{cc}%
\begin{array}
[c]{ll}%
\ \ 0 & \Lambda^{2}\\
-\Lambda^{2} & 0
\end{array}
& X_{2}\ \ \ X_{1}\\
\ast &
\end{array}
\right)  ,
\]
where $I=$\textrm{diag}$(1,1)$ and $\Lambda=$\textrm{diag}$(\lambda
_{1},\lambda_{2})$, while the bi-Hamiltonian chains (\ref{bh}) attain in these
variables the form:
\[%
\begin{array}
[c]{l}%
\pi_{0}dc_{1}=0\\
\pi_{0}dh_{1}=X_{1}=\pi_{1}dc_{1}\\
\pi_{0}dh_{2}=X_{2}=\pi_{1}dh_{1}\\
\ \ \ \ \ \ \ \ \ \ \ \ \ 0=\pi_{1}dh_{2}%
\end{array}
\ \ \ \ \
\begin{array}
[c]{l}%
\pi_{0}dc_{1}=0\\
\pi_{0}dh_{1}=X_{1}=\pi_{2}dc_{2}\\
\pi_{0}dh_{2}=X_{2}=\pi_{2}dc_{1}\\
\ \ \ \ \ \ \ \ \ \ \ \ \ 0=\pi_{2}dh_{1}%
\end{array}
\ \ \ \ \
\begin{array}
[c]{l}%
\pi_{1}dc_{2}=0\\
\pi_{1}dc_{1}=X_{1}=\pi_{2}dc_{2}\\
\pi_{1}dh_{1}=X_{2}=\pi_{2}dc_{1}\\
\ \ \ \ \ \ \ \ \ \ \ \ \ 0=\pi_{2}dh_{1}%
\end{array}
\ .
\]
For $s=1$ this system is represented by the separation curve
\begin{equation}
\bar{\lambda}^{3}+\bar{c}_{2}\lambda^{3}+\bar{c}_{1}\bar{\lambda}^{-1}+\bar
{h}_{1}\bar{\lambda}+\bar{h}_{2}=\bar{\lambda}\bar{\mu}^{2} \label{k2}%
\end{equation}
The variables $(\bar{\lambda},\bar{\mu},\bar{c})$ associated with the curve
(\ref{k2}) are connected with the variables $(\lambda,\mu,c)$ through the
Miura map (\ref{miura})\ that in this case attains the form%
\begin{equation}
\bar{\lambda}_{1}=\lambda_{1},\text{ }\bar{\lambda}_{2}=\lambda_{2},\text{
}\bar{\mu}_{1}=\lambda_{1}^{-1}\mu_{1},\text{ }\bar{\mu}_{2}=\lambda_{2}%
^{-1}\mu_{2},\text{ }\bar{c}_{1}=h_{2}(\lambda,\mu,c),\text{ }\bar{c}%
_{2}=c_{2}. \label{zm2}%
\end{equation}
In variables (\ref{zm2}) the matrix form of Poisson operators $\pi_{r}$ are%
\[
\pi_{0}(\bar{\lambda},\bar{\mu},\bar{c})=\left(
\begin{array}
[c]{cc}%
\begin{array}
[c]{ll}%
\ \ 0 & \Lambda^{-1}\\
-\Lambda^{-1} & 0
\end{array}
& X_{2}\ \ \ 0\\
\ast &
\end{array}
\right)  ,\ \ \ \pi_{1}(\bar{\lambda},\bar{\mu},\bar{c})=\left(
\begin{array}
[c]{cc}%
\begin{array}
[c]{ll}%
\ \ 0 & I\\
-I & 0
\end{array}
& 0\ \ \ 0\\
\ast &
\end{array}
\right)  ,
\]%
\[
\pi_{2}(\bar{\lambda},\bar{\mu},\bar{c})=\left(
\begin{array}
[c]{cc}%
\begin{array}
[c]{ll}%
\ \ 0 & \Lambda\\
-\Lambda & 0
\end{array}
& 0\ \ \ X_{1}\\
\ast &
\end{array}
\right)  ,
\]
(with $\Lambda=$\textrm{diag}$(\bar{\lambda}_{1},\bar{\lambda}_{2})$), while
the bi-Hamiltonian chains (\ref{bh}) are
\[%
\begin{array}
[c]{l}%
\pi_{0}d\bar{h}_{1}=0\\
\pi_{0}d\bar{h}_{2}=X_{1}=\pi_{1}d\bar{h}_{1}\\
\pi_{0}d\bar{c}_{1}=X_{2}=\pi_{1}d\bar{h}_{2}\\
\ \ \ \ \ \ \ \ \ \ \ \ \ 0=\pi_{1}d\bar{c}_{1}%
\end{array}
\ \ \ \ \
\begin{array}
[c]{l}%
\pi_{0}d\bar{h}_{1}=0\\
\pi_{0}d\bar{h}_{2}=X_{1}=\pi_{2}d\bar{c}_{2}\\
\pi_{0}d\bar{c}_{1}=X_{2}=\pi_{2}d\bar{h}_{1}\\
\ \ \ \ \ \ \ \ \ \ \ \ \ 0=\pi_{2}d\bar{h}_{2}%
\end{array}
\ \ \ \ \
\begin{array}
[c]{l}%
\pi_{1}d\bar{c}_{2}=0\\
\pi_{1}d\bar{h}_{1}=X_{1}=\pi_{2}d\bar{c}_{2}\\
\pi_{1}d\bar{h}_{2}=X_{2}=\pi_{2}d\bar{h}_{1}\\
\ \ \ \ \ \ \ \ \ \ \ \ \ 0=\pi_{2}d\bar{h}_{2}%
\end{array}
\ .
\]
Finally, for $s=2$ the system is represented by the separation curve
\begin{equation}
\tilde{\lambda}^{2}+\tilde{c}_{2}^{-1}\tilde{\lambda}+\tilde{c}_{1}^{-2}%
\tilde{\lambda}+\tilde{h}_{1}\tilde{\lambda}+\tilde{h}_{2}=\tilde{\lambda}%
^{2}\tilde{\mu}^{2}. \label{k3}%
\end{equation}
The variables $(\tilde{\lambda},\tilde{\mu},\tilde{c})$ associated with the
curve (\ref{k3}) are connected with the variables $(\lambda,\mu,c)$ of the
curve (\ref{k1}) through the Miura map (\ref{miura}), that in this case
attains the form%
\begin{equation}
\tilde{\lambda}_{1}=\lambda_{1},\text{ }\tilde{\lambda}_{2}=\lambda_{2},\text{
}\tilde{\mu}_{1}=\lambda_{1}^{-2}\mu_{1},\text{ }\tilde{\mu}_{2}=\lambda
_{2}^{-2}\mu_{2},\text{ }\tilde{c}_{1}=h_{2}(\lambda,\mu,c),\text{ }\tilde
{c}_{2}=h_{1}(\lambda,\mu,c) \label{zm3}%
\end{equation}
and they are also connected with the variables $(\bar{\lambda},\bar{\mu}%
,\bar{c})$ associated with the curve (\ref{k2}) through the Miura map
(\ref{Mrs}) that attains the form%
\[
\tilde{\lambda}_{1}=\bar{\lambda}_{1},\text{ \ }\tilde{\lambda}_{2}%
=\bar{\lambda}_{2},\text{ \ }\tilde{\mu}_{1}=\bar{\lambda}_{1}^{-1}\bar{\mu
}_{1},\text{ \ }\tilde{\mu}_{2}=\bar{\lambda}_{2}^{-1}\bar{\mu}_{2},\text{
\ }\tilde{c}_{1}=\bar{c}_{1},\text{ \ }\tilde{c}_{2}=\bar{h}_{2}(\bar{\lambda
},\bar{\mu},\bar{c}).
\]
The matrix representations of the Poisson operators $\pi_{r}$ in the variables
(\ref{zm3}) are as follows%
\[
\pi_{0}(\tilde{\lambda},\tilde{\mu},\tilde{c})=\left(
\begin{array}
[c]{cc}%
\begin{array}
[c]{ll}%
\ \ 0 & \Lambda^{-2}\\
-\Lambda^{-2} & 0
\end{array}
& X_{2}\ \ \ X_{1}\\
\ast &
\end{array}
\right)  ,\ \ \ \pi_{1}(\tilde{\lambda},\tilde{\mu},\tilde{c})=\left(
\begin{array}
[c]{cc}%
\begin{array}
[c]{ll}%
\ \ 0 & \Lambda^{-1}\\
-\Lambda^{-1} & 0
\end{array}
& 0\ \ \ X_{2}\\
\ast &
\end{array}
\right)  ,
\]%
\[
\pi_{2}(\tilde{\lambda},\tilde{\mu},\tilde{c})=\left(
\begin{array}
[c]{cc}%
\begin{array}
[c]{ll}%
\ \ 0 & I\\
-I & 0
\end{array}
& 0\ \ \ 0\\
\ast &
\end{array}
\right)  ,
\]
(with $\Lambda=$\textrm{diag}$(\tilde{\lambda}_{1},\tilde{\lambda}_{2})$)
while the bi-Hamiltonian chains (\ref{bh}) are
\[%
\begin{array}
[c]{l}%
\pi_{0}d\tilde{h}_{2}=0\\
\pi_{0}d\tilde{c}_{2}=X_{1}=\pi_{1}d\tilde{h}_{2}\\
\pi_{0}d\tilde{c}_{1}=X_{2}=\pi_{1}d\tilde{c}_{2}\\
\ \ \ \ \ \ \ \ \ \ \ \ \ 0=\pi_{1}d\tilde{c}_{1}%
\end{array}
\ \ \ \ \
\begin{array}
[c]{l}%
\pi_{0}d\tilde{h}_{2}=0\\
\pi_{0}d\tilde{c}_{2}=X_{1}=\pi_{2}d\tilde{h}_{1}\\
\pi_{0}d\tilde{c}_{1}=X_{2}=\pi_{2}d\tilde{h}_{2}\\
\ \ \ \ \ \ \ \ \ \ \ \ \ 0=\pi_{2}d\tilde{c}_{2}%
\end{array}
\ \ \ \ \
\begin{array}
[c]{l}%
\pi_{1}d\tilde{h}_{1}=0\\
\pi_{1}d\tilde{h}_{2}=X_{1}=\pi_{2}d\tilde{h}_{1}\\
\pi_{1}d\tilde{c}_{2}=X_{2}=\pi_{2}d\tilde{h}_{2}\\
\ \ \ \ \ \ \ \ \ \ \ \ \ 0=\pi_{2}d\tilde{c}_{2}%
\end{array}
\ .
\]
Thus, all three curves (\ref{k1}), (\ref{k2}) and (\ref{k3}) represent the
same St\"{a}ckel system on extended phase space $\mathcal{M}=\mathbf{R}^{6}$
in different separable coordinates, connected by respective Miura maps
(\ref{Mrs}).
\end{example}

\subsection*{Acknowledgments}

The authors would like to thank Mathematical Institute of Silesian
University in Opava for kind hospitality extended to them in the
course of their visit to Opava where a substantial part of the
present article was written. The authors are especially grateful
for useful discussions with Prof. Artur Sergyeyev.

\section*{Appendix}

\setcounter{equation}{0} \renewcommand{\theequation}{A.\arabic{equation}}

We prove here Theorem \ref{obvious}. The Jacobian $J$ of the map (\ref{minv})
is%
\[
J=\left(
\begin{array}
[c]{cc}%
\begin{array}
[c]{ccccc}%
I_{n} &  &  &  & 0_{n\times n}\\
&  &  &  & \\
s\Lambda^{s-1}M &  &  &  & \Lambda^{s}%
\end{array}
& 0_{2n\times N}\\
& \\%
\begin{array}
[c]{cc}%
\left[  \frac{\partial\bar{h}_{s-i+1}}{\partial\bar{\lambda}_{j}}\right]
_{i=1\ldots s} & \left[  \frac{\partial\bar{h}_{s-i+1}}{\partial\bar{\mu}_{j}%
}\right]  _{i=1\ldots s}\\
& \\
0_{(N-s)\times n} & 0_{(N-s)\times n}%
\end{array}
&
\begin{array}
[c]{cc}%
\left[  V_{s-i+1}^{(-s+j+1)}\right]  _{i=1\ldots s} & \left[  V_{s-i+1}%
^{(n-s+j+1)}\right]  _{i=1\ldots s}\\
& \\
0_{(N-s)\times n} & I_{N-s}%
\end{array}
\end{array}
\right)
\]
where $\Lambda=$\textrm{diag}$(\bar{\lambda}_{1},\ldots,\bar{\lambda}_{n})$,
$M=$\textrm{diag}$(\bar{\mu}_{1},\ldots,\bar{\mu}_{n})$. Thus, the vector
field $\bar{X}_{r}$ has in the coordinates $(\lambda,\mu,c)$ the components
given by%
\begin{equation}
J\left(
\begin{array}
[c]{c}%
\frac{\partial\bar{h}_{r}}{\partial\bar{\mu}}\\
-\frac{\partial\bar{h}_{r}}{\partial\bar{\lambda}}\\
0\\
0
\end{array}
\right)  =\left(
\begin{array}
[c]{c}%
\frac{\partial\bar{h}_{r}}{\partial\bar{\mu}}\\
s\Lambda^{s-1}M\frac{\partial\bar{h}_{r}}{\partial\bar{\mu}}-\Lambda^{s}%
\frac{\partial\bar{h}_{r}}{\partial\bar{\lambda}}\\
\left\{  \bar{h}_{s-i+1},\bar{h}_{r}\right\}  _{\pi_{s}}\\
0
\end{array}
\right)  \label{A1}%
\end{equation}
where $\left\{  \bar{h}_{s-i+1},\bar{h}_{r}\right\}  _{\pi_{s}}=\pi_{s}%
(d\bar{h}_{s-i+1},d\bar{h}_{r})=0$ by construction (cf. (\ref{konwencja})).
Consider now all $n$ separation relations following from the curve
(\ref{ext}):%
\begin{equation}
\sigma(\lambda_{i})+c_{N}\lambda_{i}^{n+N-1}+\ldots+c_{1}\lambda_{i}^{n}%
+h_{1}\lambda_{i}^{n-1}+\ldots+h_{n}=f(\lambda_{i})\mu_{i}^{2}\text{,
\ \ }i=1,\ldots,n \label{A1a}%
\end{equation}
These relations become identities with respect to all variables $(\lambda
,\mu,c)$ if we insert the Hamiltonians $h_{i}$ generated by (\ref{ext}) into
(\ref{A1a}). Differentiating each of these identities with respect to $\mu
_{p}$ yields%

\[
\frac{\partial h_{1}}{\partial\mu_{p}}\lambda_{i}^{n-1}+\ldots+\frac{\partial
h_{n}}{\partial\mu_{p}}=2f(\lambda_{i})\mu_{i}\delta_{ip},
\]
so that, by inverting the Vandermonde matrix $\lambda_{i}^{n-j}$ we obtain
that for any $r$ and any $p$%
\[
\frac{\partial h_{r}}{\partial\mu_{p}}=-\sum_{i=1}^{n}\frac{\partial\rho_{r}%
}{\partial\lambda_{i}}\frac{2f(\lambda_{i})\mu_{i}\delta_{ip}}{\Delta_{i}%
}=-\frac{\partial\rho_{r}}{\partial\lambda_{p}}\frac{2f(\lambda_{p})\mu_{p}%
}{\Delta_{p}}%
\]
Performing the analogous operation on separation relations
\begin{equation}
\sigma(\bar{\lambda}_{i})+\bar{c}_{N}\bar{\lambda}_{i}^{n+N-1}+\ldots+\bar
{c}_{s+1}\bar{\lambda}_{i}^{n+s}+\bar{h}_{1}\bar{\lambda}_{i}^{n+s-1}%
+\ldots+\bar{h}_{n}\lambda_{i}^{s}+\bar{c}_{s}\bar{\lambda}_{i}^{s-1}%
+\ldots+\bar{c}_{1}=f(\bar{\lambda}_{i})\bar{\lambda}_{i}^{2s}\bar{\mu}%
_{i}^{2} \label{A1b}%
\end{equation}
we obtain that for any $r$ and any $p$%
\begin{equation}
\frac{\partial\bar{h}_{r}}{\partial\bar{\mu}_{p}}=-\frac{\partial\bar{\rho
}_{r}}{\partial\bar{\lambda}_{p}}\frac{2f(\bar{\lambda}_{p})\bar{\lambda}%
_{p}^{s}\bar{\mu}_{p}}{\bar{\Delta}_{p}}\overset{(\ref{minv})}{=}%
-\frac{\partial\rho_{r}}{\partial\lambda_{p}}\frac{2f(\lambda_{p})\mu_{p}%
}{\Delta_{p}}=\frac{\partial h_{r}}{\partial\mu_{p}}. \label{A2}%
\end{equation}

Thus, due to (\ref{A1}), $\bar{X}_{r}=$ $X_{r}$ provided that $s\lambda
_{p}^{s-1}\bar{\mu}_{p}\frac{\partial\bar{h}_{r}}{\partial\bar{\mu}_{p}%
}-\lambda_{p}^{s}\frac{\partial\bar{h}_{r}}{\partial\bar{\lambda}_{p}}%
=-\frac{\partial h_{r}}{\partial\lambda_{p}}$ for all $p$ or, due to
(\ref{A2}), provided that%
\begin{equation}
\frac{\partial\bar{h}_{r}}{\partial\bar{\lambda}_{p}}=\lambda_{p}^{-s}%
\frac{\partial h_{r}}{\partial\lambda_{p}}+s\lambda_{p}^{-s-1}\mu_{p}%
\frac{\partial h_{r}}{\partial\mu_{p}}.\label{A3}%
\end{equation}
Differentiating (\ref{A1a}) with respect to $\lambda_{p}$ yields%
\[
\frac{\partial\sigma(\lambda_{i})}{\partial\lambda_{p}}\delta_{ip}+\sum
_{k=1}^{N}(n+k-1)c_{k}\delta_{ip}\lambda_{i}^{n+k-2}+\sum_{k=1}^{n}%
\frac{\partial h_{k}}{\partial\lambda_{p}}\lambda_{i}^{n-k}+\sum_{k=1}%
^{n}(n-k)\delta_{ip}h_{k}\lambda_{i}^{n-k-1}=\frac{\partial f(\lambda_{i}%
)}{\partial\lambda_{p}}\delta_{ip}\mu_{i}^{2},\ \ \ \ i=1,\ldots,n.
\]
So, for any $r$ and any $p$%
\begin{align*}
\frac{\partial h_{r}}{\partial\lambda_{p}} &  =-\sum_{i=1}^{n}\frac
{\partial\rho_{r}}{\partial\lambda_{i}}\frac{\delta_{ip}}{\Delta_{i}}\left(
\frac{\partial f(\lambda_{i})}{\partial\lambda_{p}}\mu_{i}^{2}-\frac
{\partial\sigma(\lambda_{i})}{\partial\lambda_{p}}-\sum_{k=1}^{N}%
(n+k-1)c_{k}\lambda_{i}^{n+k-2}-\sum_{k=1}^{n}(n-k)h_{k}\lambda_{i}%
^{n-k-1}\right)  \\
&  =-\frac{\partial\rho_{r}}{\partial\lambda_{p}}\frac{1}{\Delta_{p}}\left(
\frac{\partial f(\lambda_{p})}{\partial\lambda_{p}}\mu_{p}^{2}-\frac
{\partial\sigma(\lambda_{p})}{\partial\lambda_{p}}-\sum_{k=1}^{N}%
(n+k-1)c_{k}\lambda_{p}^{n+k-2}-\sum_{k=1}^{n}(n-k)h_{k}\lambda_{p}%
^{n-k-1}\right)  .
\end{align*}
Performing the analogous operation on separation relations (\ref{A1b}) and
applying Miura map (\ref{minv}) we find, after some calculation, that for any
$r$, $p$ and $i$
\[
\frac{\partial\sigma(\lambda_{i})}{\partial\lambda_{p}}\delta_{ip}+\sum
_{k=1}^{N}(n+k-1)c_{k}\delta_{ip}\lambda_{i}^{n+k-2}+\sum_{k=1}^{n}%
\frac{\partial\bar{h}_{r}}{\partial\bar{\lambda}_{p}}\lambda_{i}^{n+s-k}%
+\sum_{k=1}^{n}(n-k)\delta_{ip}h_{k}\lambda_{i}^{n-k-1}=\left(  \frac{\partial
f(\lambda_{i})}{\partial\lambda_{p}}+2sf(\lambda_{i})\lambda_{i}^{-1}\right)
\delta_{ip}\mu_{i}^{2}.
\]
Hence,%
\begin{align*}
\frac{\partial\bar{h}_{r}}{\partial\bar{\lambda}_{p}} &  =-\sum_{i=1}^{n}%
\frac{\partial\rho_{r}}{\partial\lambda_{i}}\frac{\lambda_{i}^{-s}\delta_{ip}%
}{\Delta_{i}}\left(  \left(  \frac{\partial f(\lambda_{i})}{\partial
\lambda_{p}}+2sf(\lambda_{i})\lambda_{i}^{-1}\right)  \mu_{i}^{2}%
-\frac{\partial\sigma(\lambda_{i})}{\partial\lambda_{p}}-\sum_{k=1}%
^{N}(n+k-1)c_{k}\lambda_{i}^{n+k-2}-\sum_{k=1}^{n}(n-k)h_{k}\lambda
_{i}^{n-k-1}\right)  \\
&  =-\frac{\partial\rho_{r}}{\partial\lambda_{p}}\frac{\lambda_{p}^{-s}%
}{\Delta_{p}}\left(  \left(  \frac{\partial f(\lambda_{p})}{\partial
\lambda_{p}}+2sf(\lambda_{p})\lambda_{p}^{-1}\right)  \mu_{p}^{2}%
-\frac{\partial\sigma(\lambda_{p})}{\partial\lambda_{p}}-\sum_{k=1}%
^{N}(n+k-1)c_{k}\lambda_{p}^{n+k-2}-\sum_{k=1}^{n}(n-k)h_{k}\lambda
_{p}^{n-k-1}\right)  \\
&  =\lambda_{p}^{-s}\frac{\partial h_{r}}{\partial\lambda_{p}}-\frac
{\partial\rho_{r}}{\partial\lambda_{p}}\frac{\lambda_{p}^{-s}}{\Delta_{p}%
}2sf(\lambda_{p})\lambda_{p}^{-1}\mu_{p}^{2}=\lambda_{p}^{-s}\frac{\partial
h_{r}}{\partial\lambda_{p}}+s\lambda_{p}^{-s-1}\mu_{p}\frac{\partial h_{r}%
}{\partial\mu_{p}},
\end{align*}
that is exactly the condition (\ref{A3}).

\end{document}